\documentclass[journal=ancac3,manuscript=article]{achemso}
\usepackage{textcomp,gensymb,amsmath,xcolor}
\usepackage[T1]{fontenc}
\usepackage[normalem]{ulem}
\usepackage{pdfpages}
\newcommand*\2{$_2$}

\author{Nithin Abraham}
\affiliation[IISC]
{Department of Electrical Communication Engineering, Indian Institute of Science, Bangalore 560012, India}
\author{Krishna Murali}
\affiliation[IISC]
{Department of Electrical Communication Engineering, Indian Institute of Science, Bangalore 560012, India}
\author{Kenji Watanabe}
\affiliation[NIMS1]
{Research Center for Functional Materials, National Institute for Materials Science, 1-1 Namiki, Tsukuba 305-0044, Japan}
\author{Takashi Taniguchi}
\affiliation[NIMS2]
{International Center for Materials Nanoarchitectonics, National Institute for Materials Science,  1-1 Namiki, Tsukuba 305-0044, Japan}
\author{Kausik Majumdar}
\email{kausikm@iisc.ac.in}
\affiliation[IISC]
{Department of Electrical Communication Engineering, Indian Institute of Science, Bangalore 560012, India}

\title{Astability \textit{versus} Bistability in van der Waals Tunnel Diode for Voltage Controlled Oscillator and Memory Applications}

\AtEndDocument{\includepdf[pages={2-5}]{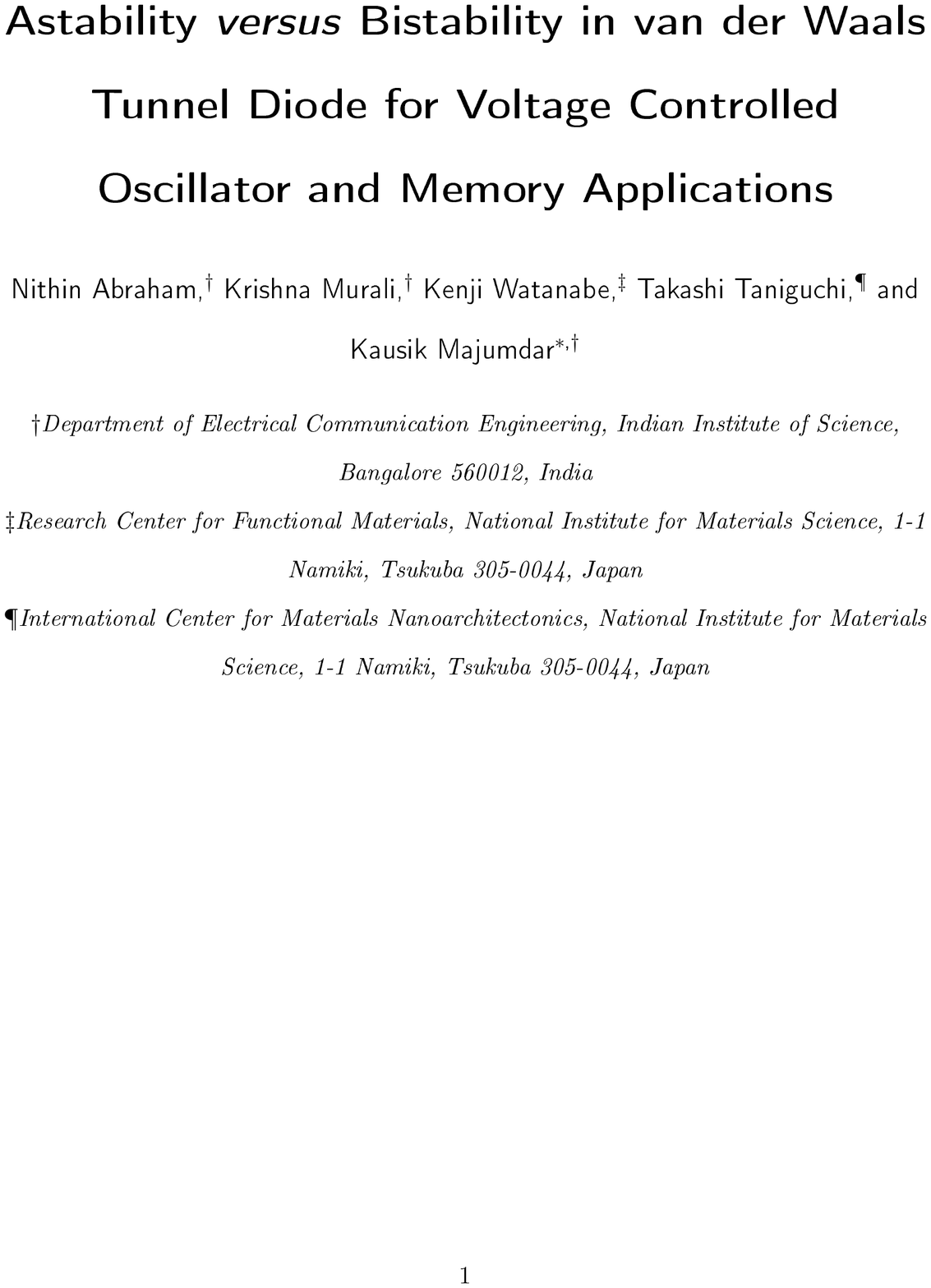}}
\begin{document}

\begin{abstract}
Van der Waals (vdW) tunnel junctions are attractive due to their atomically sharp interface, gate tunablity, and robustness against lattice mismatch between the successive layers. However, the negative differential resistance (NDR) demonstrated in this class of tunnel diodes often exhibits noisy behaviour with low peak current density, and lacks robustness and repeatability, limiting their practical circuit applications. Here we propose a strategy of using a 1L-WS\2 as an optimum tunnel barrier sandwiched in a broken gap tunnel junction of highly doped black phosphorus (BP) and SnSe\2. We achieve high yield tunnel diodes exhibiting highly repeatable, ultra-clean, and gate tunable NDR characteristics with a signature of intrinsic oscillation, and a large peak-to-valley current ratio (PVCR) of 3.6 at 300 K (4.6 at 7 K), making them suitable for practical applications. We show that the thermodynamic stability of the vdW tunnel diode circuit can be tuned from astability to bistability by altering the constraint through choosing a voltage or a current bias, respectively. In the astable mode under voltage bias, we demonstrate a compact, voltage controlled oscillator without the need for an external tank circuit. In the bistable mode under current bias, we demonstrate a highly scalable, single element one-bit memory cell that is promising for dense random access memory applications in memory intensive computation architectures.
\end{abstract}
\newpage

Nonlinear electronic devices play a pivotal role in wide ranging applications including oscillators, amplifiers, switching elements\cite{Feiginov2011,Peng2012}, and more recently in neural networks\cite{Tanaka2018,Pickett2013}. Devices exhibiting negative differential resistance (NDR) characteristics, where a negative correlation exists between device current and voltage, are excellent candidates for these interesting applications. Following the landmark discovery by Esaki\cite{Esaki1958}, tunnel diodes employing band-to-band tunneling (BTBT) have been an active area of research due to their NDR behaviour and ultra-fast response owing to the tunnelling nature of the carrier transport. Maintaining a sharp junction with a steep doping gradient is a key factor in achieving high tunneling efficiency in a tunnel diode. In this context, two-dimensional layered materials and their van der Waals (vdW) heterojunctions are highly promising\cite{Shim2016,Fan2019,Nourbakhsh2016,Movva2016,Duong2018,Jia2017,Srivastava2019,Fan2020,Roy2015,Lee2020,Roy2016,Yan2015} due to their atomically sharp junctions, which can substantially reduce the process complexity and cost as opposed to bulk semiconductors. The wide collection of layered materials allows us to choose the desirable band offset in a vdW heterojunction without worrying about the lattice mismatch between two successive layers\cite{Geim2013,Liu2016}, providing a tremendous advantage over bulk semiconductors. In addition, tunnel diodes with the ultra-thin layers provide strong gate tunability\cite{Shim2016,Fan2019,Nourbakhsh2016,Movva2016,Lee2020,Duong2018,Jia2017,Srivastava2019,Fan2020,Roy2015} - a trait which is usually unavailable in conventional tunnel diodes.

However, in a vdW heterojunction, the absence of a conventional depletion region due to ultra-thin nature of the layers, and atomic sharpness of the interface, prevents a voltage drop at the tunnel junction which degrades the NDR characteristics. This is because, for NDR to be observable, the tunneling current through the \textit{p-n} junction must be tunable which requires a voltage modulation of the spectral overlap between the filled states at the conduction band of \textit{n-}side and the empty states at the valence band of \textit{p-}side. One thus requires the presence of a spatial gap between the \textit{n} and \textit{p} regions created during the transfer process or unintentional oxidation or a tunnel barrier that acts like an artificial depletion region to accommodate the required voltage drop at the tunnel junction. Relying on these processes takes a toll on the device yield, and the achieved NDR characteristics are inferior compared to state of the art Si and III-V semiconductor based tunnel diodes\cite{Pawlik2012,Oehme2009}, in terms of limited peak current density, poor repeatability, and large noise present in the tunneling current. This has led to limited efforts to use vdW heterojunction based tunnel diodes in real life appliations\cite{Shim2016,Nourbakhsh2016}, and thus non-Esaki vdW devices are being explored to achieve NDR\cite{Mahajan2020}.

In this work, by optimizing the tunnel barrier layer and using a clean device fabrication technique that maintains the high quality of the interfaces, we demonstrate a SnSe\2/1L-WS\2/BP vdW heterostructure based broken gap tunnel diode that exhibits ultra-clean, highly repeatable, and gate tunable NDR characteristics with a large peak-to-valley ratio (PVCR). We also show that the thermodynamic stability of the tunnel diode circuit in the NDR regime can be switched by changing the external constraint from voltage bias to a current bias. This allows us to further demonstrate a reactive element free persistent oscillation in an astable mode and a highly scalable memory cell in a bistable mode of operation of the vdW heterojunction tunnel diode.

\section{Results and Discussions}
\begin{figure}[h]
  \centering
  \includegraphics[width=15.23cm]{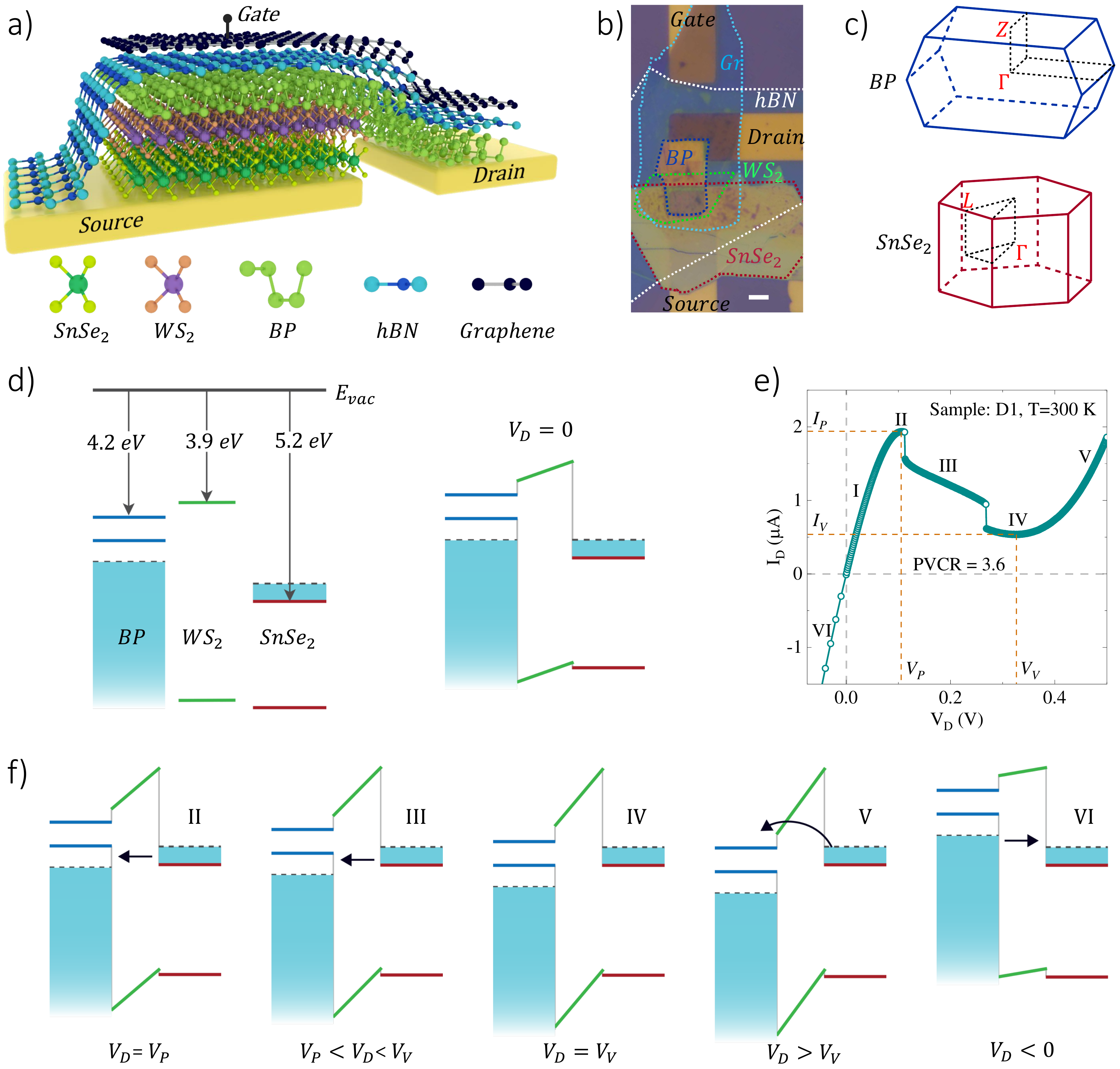}
  \caption{\label{fig:schema}\textbf{Tunnel diode operation:} \textbf{a)} Schematic illustration of the SnSe\2/1L-WS\2/BP tunnel diode with an encapsulating hBN layer and a graphene gate. \textbf{b)} A representative optical image of the heterostructure outlining the various layers. Scale bar is $5~\mu$m. \textbf{c)} Brillouin zones for bulk BP (top) and bulk SnSe\2(bottom). $Z$ and $L$ valleys correspond to valance band maxima of BP and conduction band minima of SnSe\2, respectively. \textbf{d)} Energy band alignment of the layers before (left) and after (right) contacting, suggesting broken gap alignment between BP and SnSe\2 with an ultra-thin 1L-WS\2 tunnel barrier. \textbf{e)} Measured output characteristics of the tunnel diode (D1) exhibiting a PVCR of $3.6$ at $300$ K. Various regions of operation under forward bias (Esaki mode) are marked as I through V, while the reverse bias (Zener mode) is shown as region VI. \textbf{f)} Energy band alignment for different regions of operation (see text for details). Black arrows indicate electron transport direction.}
\end{figure}

Architecture of the SnSe\2/1L-WS\2/BP tunnel diode is schematically illustrated in Figure \ref{fig:schema}a. Mechanically exfoliated multi-layer SnSe\2 flakes are identified by optical contrast and are dry transferred on to pre-patterned Au electrodes. 1L-WS\2 and multi-layer BP flakes are stacked over the SnSe\2 flake with precise alignment following the same method. The other end of the BP layer is in contact with another Au electrode. The entire stack is immediately capped with a few-layer-thick hBN flake. The hBN flake serves the dual role of a protective capping for the BP flake\cite{Doganov2015,Avsar2015} and that of a gate dielectric. The top gate contact is defined using a few-layer graphene flake which is connected to a third Au electrode. The flakes are characterized using Raman spectroscopy as shown in \textbf{Supplementary Figure 1}. The fabrication approach we employ here utilising pre-patterned contacts does not require any chemical processing step after the flake transfer, and thus preserves the pristine nature of the BP interface. More details of the fabrication process are given in \textbf{Methods} and illustrated stepwise in \textbf{Supplementary Figure 2}. A representative optical image of the stack is shown in Figure \ref{fig:schema}b. The Au electrode contacting the SnSe\2 (BP) flake is referred to as the source (drain) contact throughout the text. We measure the drain current ($I_D$) by applying an external voltage ($V_D$) at the drain contact, keeping the source terminal grounded. Seven such tunnel diodes (D1-D7) are fabricated, and all the devices exhibit repeatable NDR characteristics (summarized in \textbf{Supplementary Table 1}), pointing to the high yield of the fabrication process.

Band extrema of interest occur at the $Z$ valley for BP\cite{Qiao2014} valance band and at the $L$ valley for SnSe\2\cite{Gonzalez2016} conduction band as shown in Figure \ref{fig:schema}c. The presence of vacancies and impurities induce a large \textit{p-}type doping in BP\cite{Kiraly2017,Gaberle2018} and a large \textit{n-}type doping in SnSe\2\cite{Yan2015}. The degenerate doping and large band offsets between BP\cite{Shim2016} and SnSe\2\cite{Fan2019,Aretouli2016,Krishna2017,Murali2018} result in a type-III (broken gap) band alignment at the heterojunction\cite{Yan2015} (Figure \ref{fig:schema}d). Under forward bias, the electrons experience inelastic band-to-band tunneling (BTBT) from the filled states of SnSe\2 conduction band to the empty states of BP valence band through the monolayer-thick WS\2 tunnel barrier. The current - voltage ($I_D$-$V_D$) characteristics of the tunnel junction D1 at $300$ K is shown in Figure \ref{fig:schema}e, indicating a large current under reverse bias (Zener mode) and conspicuous NDR characteristics under forward bias (Esaki mode) - a direct evidence of BTBT dominated carrier transport\cite{Esaki1958}. The band alignment in the heterojunction under different regions of operation is depicted in Figure \ref{fig:schema}f. Under forward bias, $I_D$ increases linearly with $V_D$ at low bias (region I) to a ``peak current'' $I_P$ (point II, at $V_D=V_P$), which is attributed to an increasing overlap between filled states in SnSe\2 conduction band and empty states in BP valence band. With further increase in $V_D$, $I_D$ drops abruptly, and followed by a gradual reduction (region III) and another abrupt drop and then a slow decrease to a minimum ``valley current'' $I_V$ (region IV, at $V_D=V_V$). Such decrease in $I_D$ with an increase in $V_D$ in between the peak and the valley points indicates the presence of NDR in the DC characteristics. This decrease in $I_D$ is due to a reduction in the overlapping states beyond $V_D=V_P$. Beyond the valley point, $I_D$ again increases with further increase in $V_D$, due to an enhancement in the thermionic current over the BP/SnSe\2 barrier. The diode achieves a high PVCR ($\frac{I_P}{I_V}$) of $3.6$ at 300 K. The abrupt drops in the $I_D$ characteristics has not been observed to date in vdW tunnel diodes \cite{Shim2016,Fan2019,Yan2015,Nourbakhsh2016,Movva2016,Lee2020,Duong2018,Jia2017,Srivastava2019,Roy2016,Fan2020,Roy2015}, and indicate the presence of oscillations in the diode \cite{Seabaugh2003} which will be explained later.

As explained earlier, BP and SnSe\2 form a broken gap vdW heterojunction, which is ideally suited for high BTBT current. However in the absence of the 1L-WS\2 tunnel barrier, due to the high conductivity of such junction along with its atomic sharpness, there is a very little voltage drop across the junction. This forces the quasi-Fermi levels of the BP and the SnSe\2 sides to be almost aligned at the tunneling interface in spite of a change in the applied external bias. This degrades the PVCR, with a possibility of complete suppression of NDR characteristics, although the total tunneling current remains high. When we introduce the WS\2 sandwich layer, it acts like an atomically sharp depletion layer allowing a larger voltage to drop across it. This effectively de-pins the BP and SnSe\2 bands at the tunneling interface, allowing their relative movement which is crucial in observing the NDR characteristics \cite{Yan2015,Pawlik2012}. Clearly, it is critical to optimize the thickness of the barrier layer to achieve the optimum device performance. The results of tunnel diodes fabricated with different barrier layers [no barrier layer (D8), 1L-WS\2 (D1), 2L-WS\2 (D9), and 1L-MoS\2 (D10)] are summarized in \textbf{Supplementary Table 2}. We observe that the device with no barrier exhibits an order of magnitude higher tunneling current, however, with no NDR signature (\textbf{Supplementary Figure 3a}). The device with 2L-WS\2 as a barrier layer (\textbf{Supplementary Figure 3c}) exhibits a PVCR of $\sim 2$, but with significantly suppressed tunneling current due to increased barrier width. An annealing step in the fabrication also helps in reducing the inter-layer spacing. As can be seen from samples D2 and D3 (\textbf{Supplementary Table 1}), the current density increases drastically with an annealing step after WS\2 transfer with no degradation of PVCR. The tunnelling current for 1L-MoS\2 (\textbf{Supplementary Figure 3d}) and 1L-WS\2 barrier devices are comparable, however, the NDR characteristics are far superior and cleaner in the latter case, likely due to higher defect density in MoS\2\cite{Addou2015,Bampoulis2017}.

\begin{figure}[h]
  \centering
  \includegraphics[width=11.05cm]{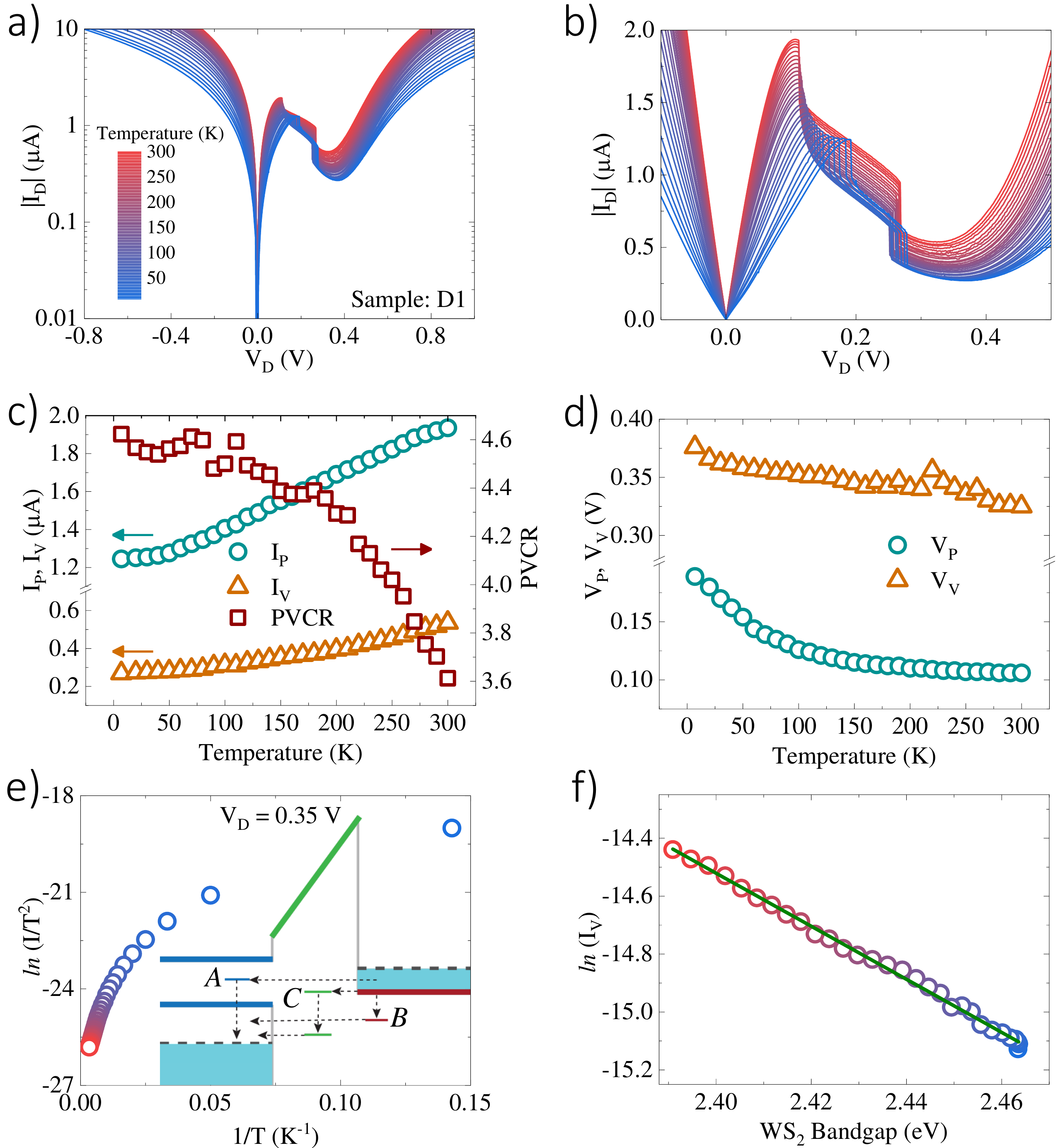}
  \caption{\label{fig:temp}\textbf{Temperature dependence of tunnel diode characteristics:} \textbf{a)} $I_D-V_D$ curves from sample D1 at temperatures ranging from $7$ to $300$ K in log scale. \textbf{b)} Enlarged view of the NDR region in linear scale. \textbf{c)} Left axis: Extracted peak (circle markers) and valley (triangle markers) currents \textit{versus} operating temperature.  The strong temperature dependence for $I_P$ suggests a phonon assisted tunnelling transport. Right axis: PVCR (square markers) of the device at each operating temperature. \textbf{d)} Variation of peak (circle markers) and valley (triangle markers) positions with temperature. \textbf{e)} Richardson plot for $I_D$ near valley ($@V_D =0.35$ V) exhibiting a positive slope indicating an origin of valley current different from thermionic process. Color of the markers are mapped to the measurement temperature. Inset: various transport pathways through gap states for the generation of excess current. \textbf{f)} Variation of $I_V$ with a temperature induced bandgap change of WS\2 (Circle markers with color mapped to temperature.). Good agreement with a the linear fit (green trace) suggest defect induced origin of excess current.}
\end{figure}

To investigate the current transport mechanisms at different regions of operation, we measure the $I_D$-$V_D$ characteristics of the device D1 at various temperatures ranging from $7$ K to $300$ K, as shown in Figure \ref{fig:temp}a. A magnified view of the NDR region in linear scale is given in Figure \ref{fig:temp}b. The NDR characteristics and the low noise nature is maintained in the entire temperature range pointing to the stability of the processes and the high quality of the interfaces. The strong temperature dependence of the peak current points to a phonon-assisted inelastic tunneling process\cite{Chynoweth1963,Seabaugh2000}. This is in agreement with the fact that conduction band minimum at SnSe\2 being at the $L$ point, while the valence band maximum of BP is at the $Z$ point (see Figure \ref{fig:schema}d), requiring the assistance of phonon for in-plane momentum conservation during the tunnelling process\cite{Kane1961}.

Extracted temperature dependence of the peak and the valley currents and the corresponding voltages are given respectively in Figure \ref{fig:temp}c and d. Figure \ref{fig:temp}c shows that with a reduction in temperature, $I_V$ reduces at a slightly faster rate than $I_P$, improving the PVCR to $\sim 4.6$ at low temperature. However, $I_V$ appears to be a much weaker function of temperature than thermionic process, suggesting a non-thermionic origin of the valley current. To establish the point quantitatively, using Richardson equation\cite{Crowell1965}, we plot $ln(\frac{I_V}{T^2})$ as a function of $\frac{1}{T}$ in Figure \ref{fig:temp}e. The positive slope observed clearly suggests that the valley current is not dominated by the thermionic current over the BP conduction band barrier. Such non-thermionic current at the valley is common in tunnel diode literature \cite{Chynoweth1961,Majumdar2014}, and is usually attributed to the excess current. The excess current originates from carrier transport through the sub-bandgap states created by structural disorders and impurities, and several possible transport mechanisms are schematically illustrated in inset of Figure \ref{fig:temp}e. The carriers from either of the degenerate regions of BP or SnSe\2 can lose energy and drop to states in the otherwise forbidden band gap (paths A and B). These carriers then tunnel to the other side resulting in a nonzero valley current. The nonzero defect density present in the WS\2 barrier layer also acts as an intermediate state for the excess current (path C). Considering the bandgap ($E_G$) of 1L-WS\2 as the effective energy gap the carriers need to overcome to generate the excess current, we first map the bandgap of 1L-WS\2 at different measurement temperatures \cite{Nagler2018}, and then plot $\ln(I_V)$ as a function of the corresponding $E_G$ in Figure \ref{fig:temp}f. The linear trend with negative slope provides further evidence of the defect induced excess current as the primary source of valley current \cite{Chynoweth1961}. The non-thermionic origin of $I_V$ sets a lower bound on the valley current and hence causes the PVCR to saturate at low temperature, as shown in right axis of Figure \ref{fig:temp}c. The suppression of the excess current by reducing the defect density through the usage of higher quality flakes is thus of paramount importance to further reduce $I_V$ and achieve a PVCR limit determined by the thermionic current.

In the presence of nonzero series resistance, the external bias required to achieve peak or valley configurations is given by
\begin{equation}\label{eq:Vpv}
V_{P,V}(T)=V_{P,V}^0+I_{P,V}(T)\times R_s(T)
\end{equation}
where $V_{P,V}^0$ is the drop at the junction at peak or valley, and $R_s(T)$ is the temperature dependent series resistance. The measured $V_P$ and $V_V$ are shown in Figure \ref{fig:temp}d as a function of the temperature. The increase in $V_P$ and $V_V$ with a decrease in temperature results from an increase in $R_s$ due to a lower injection efficiency at the contact.

\begin{figure}[h]
  \centering
  \includegraphics[width=16.6cm]{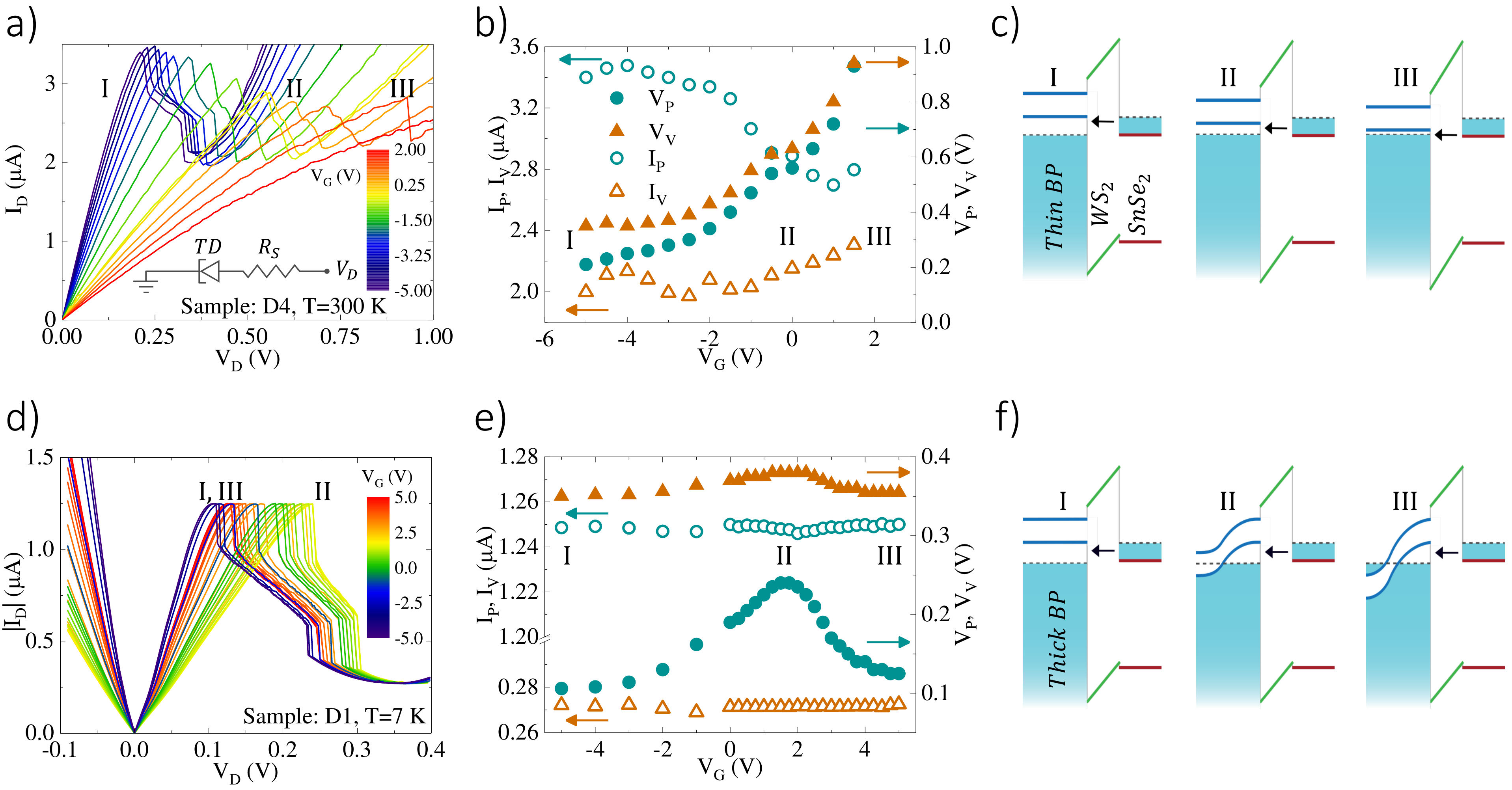}
  \caption{\label{fig:gate}\textbf{Gate modulation of tunnel diode characteristics for thin and thick BP layers:} \textbf{a)} $I_D-V_D$ curves from sample D4 employing a thin BP layer at $300$ K as a function of gate bias ranging from $2$ to $-5$ V. Inset: An equivalent circuit of the heterostructure showing the series resistance of the lateral channel $R_s$ and the tunnel junction. \textbf{b)} Left axis: Modulation of peak (empty teal circle markers) and valley (empty orange triangle markers) currents with gate bias. Increase in $I_P$ is consistent with an increasing \textit{p-}doping in BP at negative $V_G$. $I_V$ being originating from excess current exhibit a weaker dependence on $V_G$. Right axis: Peak (solid teal circle markers) and valley (solid orange triangle markers) positions occurring at larger voltages with an increasing drop across $R_s$ towards increasing $V_G$. \textbf{c)} Band alignment corresponding to $I_P$ at negative $V_G$ (I), near-zero $V_G$ (II) and positive $V_G$ (III). \textbf{d} $I_D-V_D$ curves from sample D1 employing a thick BP layer at $7$ K as a function of gate bias ranging from $5$ to $-5$ V. \textbf{e)} Extracted peak and valley currents (left axis) and positions (right axis) from D1. Invariance of $I_P$ with $V_G$ suggests a fixed \textit{p-}doping at the tunnelling interface. $V_P$ and $V_V$ gets modulated according to the variation of $R_s$ with $V_G$. \textbf{f)} Band alignment at peak configuration corresponding to I) increased doping at the top interface of the BP channel at negative $V_G$, II) maximum channel resistance at slightly positive gate bias due to the depletion region formed near top of BP, III) e-h bilayer formation at the BP layer at further positive $V_G$ and a subsequent reduction of $R_s$.}
\end{figure}

We next measure the characteristics of the tunnel diode by varying $V_G$ at a fixed temperature. $V_G$ modulates the device response through two different ways: 1) it changes the extent of the \textit{p-}doping of the BP layer at the junction and hence modulates the tunneling rate, 2) it tunes the lateral resistance $R_s$ of the BP channel. For a thin BP layer, both these effects are significant. The response of one such device (sample D4, less than $10$ nm thick BP) at 300 K is shown in Figure \ref{fig:gate}a. Modulation of $I_P$ and $I_V$ with $V_G$ is given in the left axis of Figure \ref{fig:gate}b. $I_P$ increases with more negative $V_G$ due to an enhancement in the \textit{p-}doping of the BP layer, as illustrated using the band diagram I in Figure \ref{fig:gate}c. An increased \textit{p-}doping at the tunneling interface results in a larger overlap between the filled states in the conduction band of SnSe\2 and the empty states of BP valance band and thus a larger $I_P$. Similarly, an increasingly positive $V_G$ results in a reduced \textit{p-}doping in BP as shown in Figure \ref{fig:gate}c (II and III), and hence lowers $I_P$. $I_V$ being dominated by the excess current remains largely independent of $V_G$ as expected. The modulation of the doping also affects the lateral resistance $R_s$ of the device and causes the peak and valley to occur at a larger voltage with an increase in positive $V_G$ (right axis of Figure \ref{fig:gate}b). A similar observation is made for the device response at $4.7$ K given in \textbf{Supplementary Figure 4}.

To decouple the effects of tunneling rate and series resistance, we measure the response from a device employing a thick ($\sim 20$ nm) BP layer (sample D1) at $7$ K, and the results are summarized in Figure \ref{fig:gate}d-f. $I_P$ and $I_V$ and hence the PVCR remain independent of $V_G$, as shown in left axis of Figure \ref{fig:gate}e. $I_P$ being limited by the overlap of the available states for tunneling, its invariance with $V_G$ suggests that the local hole density near the tunneling interface in the BP layer remains independent of $V_G$ due to screening\cite{Li2014}. The shift of $V_P$ and $V_V$ with varying $V_G$ (right axis of Figure \ref{fig:gate}e) without any change in $I_P$ and $I_V$ indicates a change in the series resistance with $V_G$. As $V_G$ becomes more negative, the BP channel becomes more \textit{p-}doped (region I). This reduces $R_s$ and hence the external bias required to align the junction for peak current drops, as seen in Figure \ref{fig:gate}e. On the other hand, a small positive $V_G$ (region II) reduces the \textit{p-}type doping in the BP channel and hence increases the lateral resistance leading to an increase in $V_P$. This trend follows till $V_D \approx 1.5$ V. For larger positive $V_G$ (region III), the trend turns around and $V_P$ starts decreasing. Due to the ambipolar nature of BP, at large positive $V_G$, the top gate interface of the BP channel turns electron rich, while a hole rich channel near the bottom tunneling interface is retained. This leads to the formation of an electron-hole (e-h) bilayer along the vertical direction \cite{Wu2019}, as schematically shown in Figure \ref{fig:gate}f. In the e-h bilayer, both the layers act as parallel channels of transport, but employing different types of carriers. The carrier can switch its nature by tunnelling across the barriers formed between \textit{n}$^+$ and \textit{p}$^+$ regions of BP. This process occurs efficiently due to the small and direct nature of BP bandgap \cite{Xiong2020}. This effect reduces the lateral resistance at higher positive $V_G$ and reduces $V_P$. Due to the smaller magnitude of $I_V$, the relative change in the potential drop across the series resistance is less at the valley position, resulting in a weaker gate modulation of $V_V$.

The different rates of modulation for peak and valley positions lead to a gate dependent NDR window, importance of which will be explained later in a memory device. This also tunes the slope of the current in the NDR region and hence results in a gate controllable ac resistance. Such a modulation of the ac resistance is imperative in NDR based amplifier circuits\cite{Wang2019} where the gain of the amplifier can be controlled by a gate bias.

\begin{figure}[h]
  \centering
  \includegraphics[width=16.43cm]{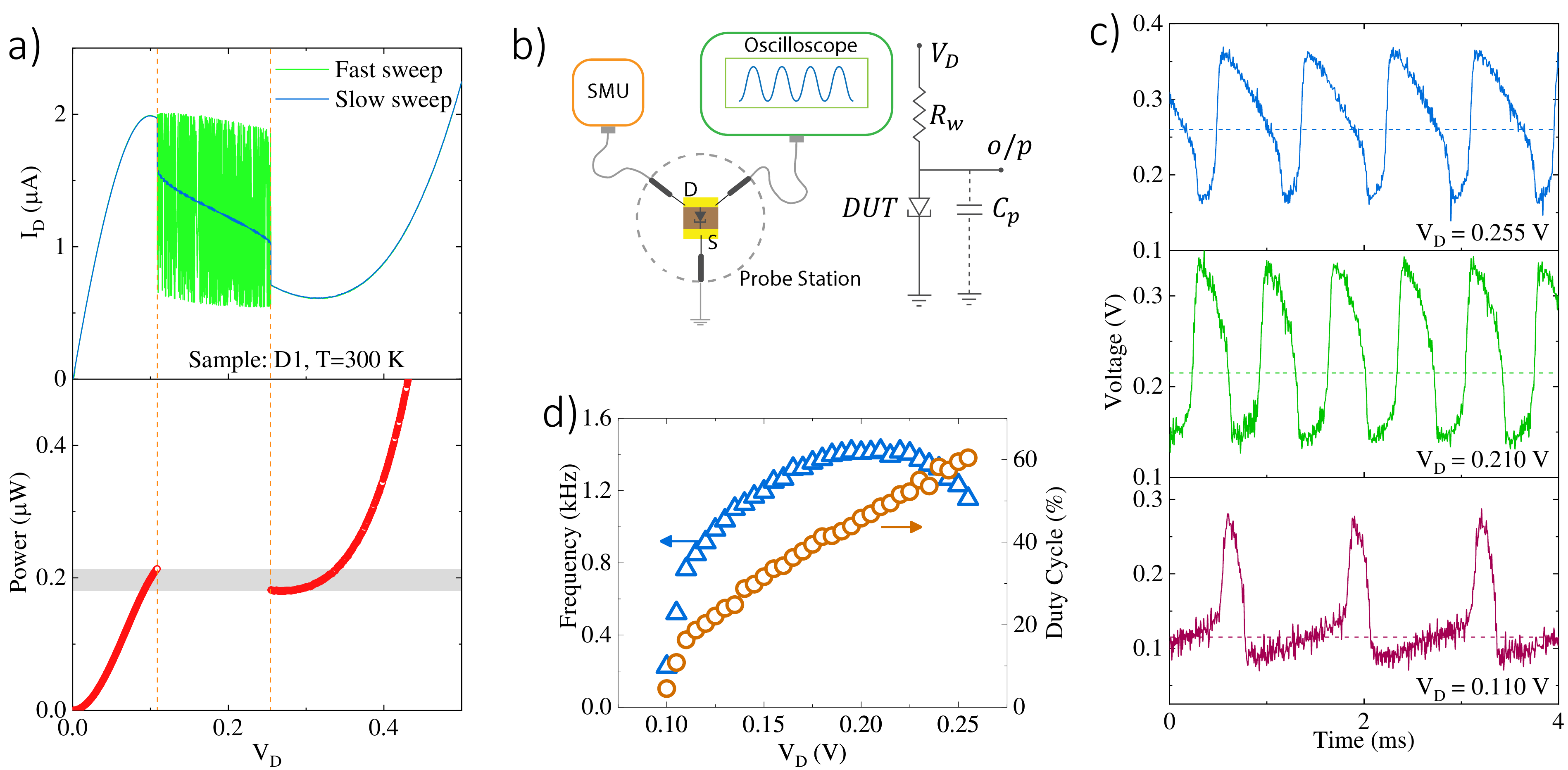}
  \caption{\label{fig:osc}\textbf{Astable operation with intrinsic oscillations:} \textbf{a)} Top: $I_D-V_D$ curves from D1 at $300$ K corresponding to slow (blue trace) and fast (green trace) $V_D$ sweep exhibiting instability at the NDR region. Bottom: Power delivered to the system at each $V_D$. The grey shaded region indicates thermodynamic instability due to multi-valued voltage for a given power delivered. \textbf{b)} Schematic of measurement setup used for observing oscillations along with an equivalent circuit showing the parasitic effects limiting the high frequency operation. \textbf{c)} Temporal response of the system at three different biases showing variation of duty cycle as well as frequency of oscillation. \textbf{d)} Measured frequency (left axis, in blue triangle markers) and duty cycle (right axis, in orange circle markers) of oscillations as a function of biasing voltage $V_D$.}
\end{figure}

NDR devices find widespread application in oscillator circuits, where the negative differential resistance of the device is used to compensate the stray resistance in the tank circuit, resulting in an undamped oscillation\cite{Mishchenko2014,Degrift1981}. However, the dependency on an external tank circuit to create oscillation is the limiting factor for large output power at high frequency \cite{Jonasson2014,Woolard1996,Zhao2003}. In the quest for intrinsic oscillator without external reactive elements, the astable nature of the NDR region is explored by measuring the output characteristics with different measurement speeds. The results of a slow (10 samples/sec) $I_D$ measurement as a function of the bias $V_D$ for D1 is given in the top panel of Figure \ref{fig:osc}a as the blue trace. On reducing the integration time for each sample point (200 samples/sec), we observe an $I_D$ oscillating between $I_P$ and $I_V$ as shown by the green trace in Figure \ref{fig:osc}a. During the slow measurement, the obtained current is a time-average of the fluctuating current. Outside this unstable region, the current measured with both configurations overlap each other perfectly.

The measured average power delivered to the entire system (illustrated in bottom panel of Figure \ref{fig:osc}a) also shows \emph{N-shape} characteristics, pointing to the thermodynamic instability in the NDR regime. This results in the multi-valued nature of the voltage output for a given power delivered. The interplay between the positive feedback in the unstable regime and the restoring force by the fixed external voltage bias leads to the observed oscillation. Note that the absence such thermodynamic instability in a device with 2L-WS\2 as the barrier rules out the possibility of sustained oscillation even though it exhibits a comparable PVCR (see \textbf{Supplementary Figure 5}).

We now bias the tunnel diode at a fixed $V_D$ in the NDR region, and measure the oscillating voltage across the diode using a digital storage oscilloscope as shown in Figure \ref{fig:osc}b \cite{Seabaugh2003}. The nonzero cabling and setup resistance $R_w$ from the tunnel diode to the SMU allows the voltage across the diode to fluctuate and show up across the oscilloscope probe. The measured temporal response at three different $V_D$ values is shown in Figure \ref{fig:osc}c, indicating stable and persistent oscillation. Such an astable operation helps in realizing single element oscillator where the need for integrating reactive elements on chip can be avoided. Proximity of the biasing point to either of peak or valley modulates the duration for which the circuit latches near that particular state, modulating the duty cycle and frequency of oscillation, as shown in Figure \ref{fig:osc}d. This can be seen as slower initial rise for a bias of $V_D=0.110$ V and a slow initial fall for $V_D=0.255$ V (bottom and top panels in Figure \ref{fig:osc}c). The system reaches a maximum frequency when biased in the middle of the NDR region as the restoring forces are maximum. The tuning of the oscillation frequency by the applied bias is shown in \textbf{Supplementary Video 1}. Note that, the large stray capacitance of the cable and the probe station ($\sim75$ pF) limits the observed frequency of oscillation, while the intrinsic frequency is much higher. A reduction in the stray capacitance in the circuit coupled with an enhancement in the peak current density would help to attain a higher frequency. These oscillations are stable over long duration as observed from different measurements (different panels in \textbf{Supplementary Figure 6}) and survive at low temperature as well, as shown in \textbf{Supplementary Figure 7}.

\begin{figure}[h]
  \centering
  \includegraphics[width=10.68cm]{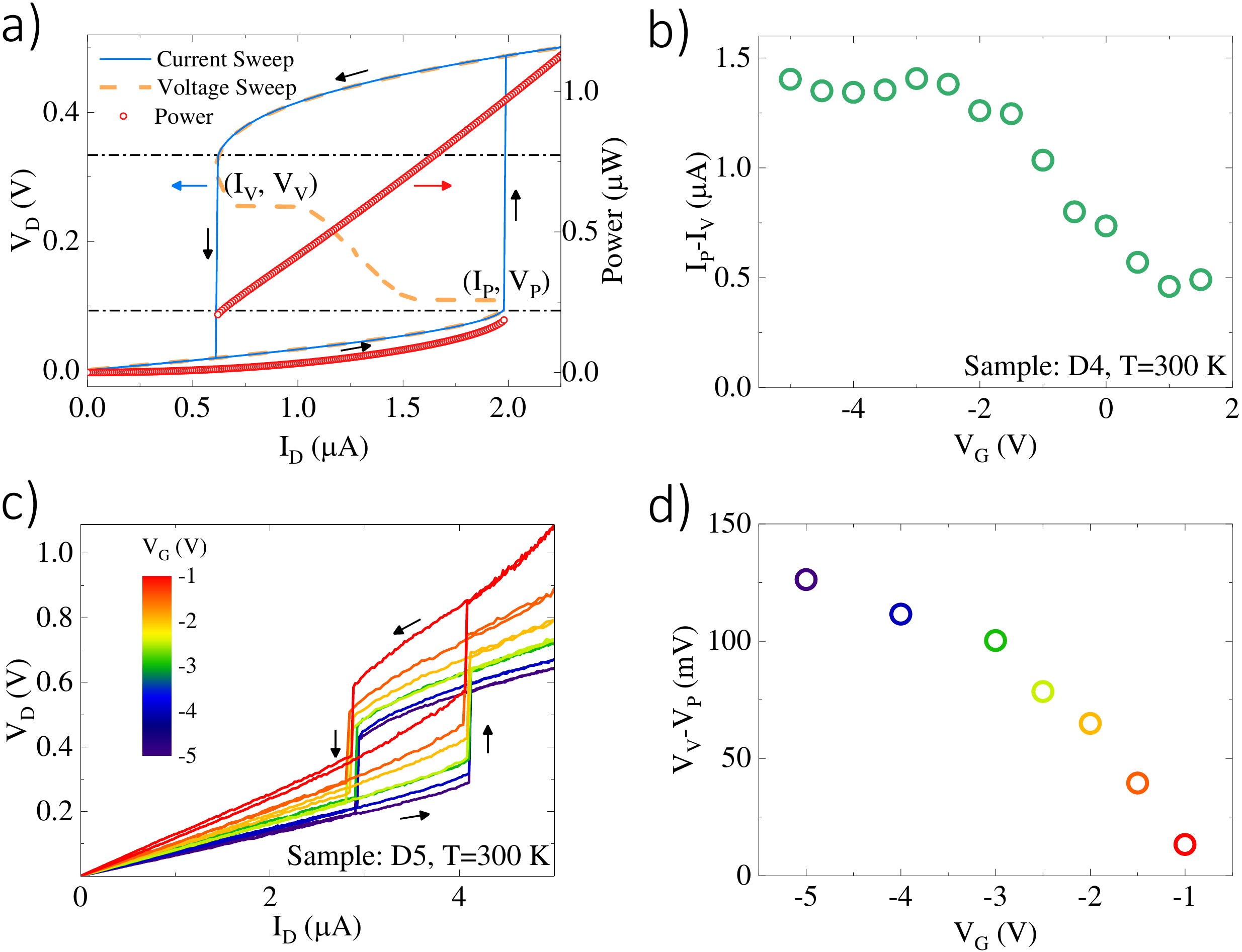}
  \caption{\label{fig:mem}\textbf{Bistable operation with tunable memory cell:} \textbf{a)} Left axis: Results from a dual current sweep (blue trace) along with a voltage sweep (orange dashed trace) as a reference. Direction of the current sweep is indicated by black arrows. A bistable window between bias of $I_P$ and $I_V$ allows storage of information. Upper (lower) branch corresponds to a stored value of logic `1' (logic `0'). Right axis: Power delivered (red circle markers) to the system as a function of the current forced through the system. Multi-valued power for a given current bias indicates presence of metastable states separated by an energy barrier. \textbf{b)} Modulation of the write window from sample D4 at $300$ K with $V_G$, suggesting a tunable noise performance. \textbf{c)} Output voltages from current sweep at $V_G$ ranging from $-1$ to $-5$ V from sample D5 with thick BP at $300$ K. \textbf{d)} Extracted read window tunability with $V_G$ (color mapped to $V_G$) showing highly versatile memory operation.}
\end{figure}

We now investigate the possibility of obtaining a bistable operation from the tunnel diode for random access memory (RAM) applications. This is achieved by forcing $I_D$ across the tunnel diode terminals and measuring the voltage drop across it, as shown in the left axis of Figure \ref{fig:mem}a (in blue trace). The forward and reverse sweeps are indicated by the black arrows. The result from a voltage sweep is also shown as dashed orange trace for reference. When the forcing current is increased beyond $I_P$, the voltage across the diode jumps to the positive differential resistance branch beyond the valley point. After this jump, when the current bias is reduced, the voltage does not trace back in the same path, rather remains in the positive resistance branch until a current bias value of $I_V$. When current bias is reduced below $I_V$, the voltage abruptly drops to the other positive differential resistance branch below the peak point. Hence the output voltage is multi-valued for a given range of current bias, and its value depends on the history of the device. This gives the device the ability to store information equivalent of one bit encoded in the voltage across it. The stored value of the junction can be altered by forcing positive or negative current spikes that ride on the biasing current. To retrieve the stored bit value, just monitoring the voltage across the device is sufficient, cutting down the extra elements required in conventional memories. A voltage value above $V_V$ and below $V_P$ corresponds to a stored value of logic `1' and logic `0' respectively (Figure \ref{fig:mem}a).

The thermodynamic origin of the bistable states can be understood from extracting the power delivered to the circuit at each current bias, as shown by the red circles in the right axis of Figure \ref{fig:mem}a. In between $I_P$ and $I_V$, at a given current bias, there are two metastable states of the system separated by an energy barrier. This is in stark contrast with the previously discussed astable oscillation at a given voltage bias. It is striking that by changing the constraint (voltage or current bias), the thermodynamic stability of the system can be effectively tuned between astability to bistability.

The logic state stored in the memory cell should be less susceptible to external noise sources and the read operation should be able to distinguish the state of the system with confidence. Thus, the threshold for switching between logics `0' and `1' is an important metric. The separation $I_P - I_V$ corresponds to the maximum allowed peak to peak fluctuation, and determines the stored bit retention. For device D4 with thin BP, $I_P$ being strongly dependent on $V_G$, results in a gate tunable $I_P-I_V$ as shown in Figure \ref{fig:mem}b. A negative gate bias is able to enhance the range by $\sim180\%$ making the memory cell more immune to external noise. On the other hand, the separation $V_{V}-V_{P}$ indicates the stability against noise during the read operation. Our ability to modulate the series resistance using $V_G$ allows this separation to be effectively tuned, as shown in Figure \ref{fig:mem}c-d for device D5 with a thick BP channel. The tunable noise margins thus obtained has a trade off with the operating frequency as well as both the standby and dynamic power consumption. This can be used in designing an intelligent memory architecture that dynamically decides the noise margin providing optimum speed and power consumption.

The proposed memory cell with a junction area of 400 nm$^2$ is expected to exhibit a standby power of $0.33$ pW (3.13 pW) when retaining logic `0' (logic `1'), which is significantly less compared with the state of the art Intel 22 nm HDC Tri-gate LP SRAM cell ($\sim 40$ pW) \cite{Jan2012}. On the other hand, the single element and vertical design of the heterojunction memory allows us to reduce the footprint per cell dramatically, providing about $100$ times higher packing density. The additional possibility of stacking multiple heterojunctions vertically using 3D integration brings about a great density advantage compared to the state of the art SRAM. This is promising for next generation architectures like memory intensive computing and bio-inspired computing.

\section{Conclusion}
In conclusion, we report a broken gap van der Waals heterojunction based tunnel diode exhibiting clean, repeatable, and gate tunable negative differential resistance characteristics with a high peak-to-valley ratio over a large temperature range. By altering the constraint from constant voltage to constant current bias, the thermodynamic stability of the tunnel diode circuit is tuned from astability to bistability. This allows us to demonstrate multi-functional operations like a reactive-element-less on chip, compact voltage controlled oscillator, and a highly scalable, single element random access memory cell for ultra-dense memory applications. The operations are sustained at very low temperature, making them attractive for cryogenic electronics as well.

\section{Methods}
\subsection{Fabrication}
Au lines are defined by optical lithography using a 360 nm UV source and AZ5214E resist spin-coated on an Si/SiO\2 substrate with 285 nm thick oxide formed by dry chlorinated thermal oxidation and forming gas annealing. A 20 nm thick Ni film followed by a 40 nm thick Au film is deposited \textit{via} DC magnetron sputtering and lifted of by Acetone/iso-propyl alcohol rinse to form the bottom contact. The SnSe\2 fakes are exfoliated from bulk crystals using Scotch tape, and are subsequently transferred to a poly-di-methyl-siloxane (PDMS) sheet. Flakes of thickness $\sim50$ nm are identified by optical contrast and transferred onto the Au line forming source contact using a dry transfer technique. This process is performed underneath a microscope using controlled translation and rotation for desired positioning of the flakes with respect to the pre-patterned substrate. This process is repeated for monolayer WS\2. The SnSe\2/WS\2 stack is annealed at 70\textdegree C for 2 min in ambient for devices D1, D2, D4 and D6. BP of thickness $\sim20$ nm is transferred next contacting another Au line forming drain contact. The stack is immediately encapsulated with hBN following the BP transfer. For devices D3 and D5, the stack is annealed after hBN transfer. Few layer graphene is transferred to form the gate contact. The layers in the whole device fabrication process do not undergo any chemical treatment and hence maintain their pristine quality and clean interfaces.
\subsection{Characterization}
The devices are loaded into a closed cycle He probe station Lakeshore CRX-6.5K or are wire bonded to a closed cycle He cryostat and characterized using a Keithley 4200A SCS parameter analyser for DC measurements and a Tektronix MDO3000 digital storage oscilloscope for frequency measurements. All the measurements were done in vacuum with a pressure $<10^{-4}$ Torr.

\section{Acknowledgements}
K. M. acknowledges the support a grant from Indian Space Research Organization (ISRO), a grant from MHRD under STARS, grants under Ramanujan Fellowship and Nano Mission from the Department of Science and Technology (DST), Government of India, and support from MHRD, MeitY and DST Nano Mission through NNetRA. K.W. and T.T. acknowledge support from the Elemental Strategy Initiative conducted by the MEXT, Japan, Grant Number JPMXP0112101001, JSPS KAKENHI Grant Numbers JP20H00354 and the CREST(JPMJCR15F3), JST.

\section{Supporting Information}
The Supporting Information is available on:
Raman characterization of flakes; Fabrication steps; Summary of results from multiple devices and different barrier layers; Gate dependence of output characteristics of sample D4 at 4.7 K; Absence of oscillation for heterostructure with thick barrier layer; Persistence and stability of oscillations; Astable operation at 7 K; Video showing modulation of oscillation frequency as a function of bias voltage.

\section{Competing Interest}
The authors declare no financial or non-financial competing interest.

\bibliography{refshort}
\newpage
\end{document}